\documentclass[conference]{IEEEtran}
\usepackage[utf8]{inputenc}
\usepackage[T1]{fontenc}
\usepackage{graphicx}
\usepackage{bm}
\usepackage{amsmath}
\usepackage{amssymb}
\usepackage{float}
\usepackage{subcaption}
\usepackage{booktabs}
\usepackage{multirow}
\usepackage{algorithm}
\usepackage{algorithmic}
\usepackage{xcolor}
\usepackage{colortbl}
\usepackage{array}
\usepackage{enumitem}
\usepackage{tikz}
\usetikzlibrary{quantikz}
\usepackage{cite}

\definecolor{bestcolor}{RGB}{144,238,144}
\definecolor{worstcolor}{RGB}{255,182,193}

\begin{document}
\title{Noise Resilience of Quantum Support Vector Machine with Selected Feature Maps}
\author{
\IEEEauthorblockN{Muhammad Ahsan Shakeel}
\IEEEauthorblockA{Department of Electrical Eng.\\Lahore University of\\Management Sciences\\Lahore 54972, Pakistan}\\ahsanshakeel13@gmail.com\and
\IEEEauthorblockN{Saad Muzammil}
\IEEEauthorblockA{Department of Electrical Eng.\\Lahore University of \\ Management Sciences,\\ Lahore 54972, Pakistan}\\saadmuzammil098@gmail.com\and
\IEEEauthorblockN{Danyal Tayyub}
\IEEEauthorblockA{Department of Electrical Eng.\\Lahore University of \\ Management Sciences,\\ Lahore 54972, Pakistan}\\danyaltayyub@gmail.com\and
\IEEEauthorblockN{Muhammad Faryad$^\ast$}
\IEEEauthorblockA{Department of Physics\\Lahore University of\\Management Sciences\\Lahore 54972, Pakistan}\\
muhammad.faryad@lums.edu.pk\\
$^\ast$Corresponding Author
}
\maketitle

\begin{abstract}
Gate-level noise degrades the classification accuracy of Quantum Support Vector Machines (QSVMs) on Noisy Intermediate-Scale Quantum (NISQ) hardware, and the degree of degradation depends on how classical data is encoded into quantum states. We tested Z, ZZ, a Pauli, and an amplitude-inspired feature maps under depolarizing, bit-flip, and phase-flip noise channels in $52$ controlled experiments with error probabilities $p =0.01, 0.05, 0.10$, and $0.50$. The amplitude-inspired feature map had $100$\% test accuracy up to $p = 0.10$ across all three noise channels, while other feature maps fell to $65$-$90$\% under the same noise level and type. The Z feature map was found to be immune to phase-flip noise to a significantly high error rate, a consequence of the commutation relation $[R_Z, Z] = 0$. Entangled circuit variants produced generalization gaps in train-tests of up to $17.5$\% under noise, whereas the amplitude variants maintained zero gap throughout. These results give practitioners data-driven criteria for a feature map on near-term quantum hardware.
\end{abstract}

\section{Introduction}

A quantum feature map encodes classical data into an $n$-qubit state, placing it in a $2^n$-dimensional Hilbert space where the inner product defines the kernel function \cite{havlicek2019supervised,rebentrost2014}. This gives quantum kernel machines representational capacity that no classical polynomial-time algorithm can exactly replicate \cite{schuld2019,huang2021}. Quantum Support Vector Machines (QSVMs) build on this idea by passing the resulting kernel matrix to a classical state vector machine (SVM) solver. Therefore, only feature mapping is performed on quantum hardware \cite{havlicek2019supervised,peters2021}, giving rise to a hybrid quantum-classical algorithm. The approach has been validated on small binary tasks using superconducting processors, but classifier accuracy under realistic noise has not been compared across competing feature maps, which is the main motivation of the work in this paper.

Noisy intermediate scale quantum (NISQ) processors carry two-qubit gate error rates between $10^{-3}$ and $10^{-2}$, coherence times on the order of microseconds, and qubit counts in the tens to low hundreds \cite{preskill2018quantum}. Each gate error shifts the affected kernel matrix entry and distorts the decision boundary that the SVM learns \cite{bharti2022}. Whether the shift is large or small depends on circuit architecture: entangling gates propagate single-qubit errors into correlated multi-qubit errors, while shallow single-qubit circuits confine damage to individual qubits. No controlled study has measured this difference across multiple feature map designs and noise channels.

Research on noise in quantum computing has focused mainly on variational algorithms, characterizing barren plateaus and vanishing gradients under depolarizing models \cite{cerezo2021,bharti2022}. Error mitigation methods such as zero-noise extrapolation and probabilistic error cancellation reduce bias at the cost of substantially more circuit executions \cite{temme2017, goto2021}. Peters et al. \cite{peters2021} measured QSVM accuracy on noisy hardware but evaluated only one feature map design. No comparative study spans multiple kernel architectures and noise channels simultaneously.

This paper closes that gap through a controlled study covering four feature map architectures, three noise channels, and four error probability levels ($52$ experiments total), making four specific contributions:

\begin{enumerate}[leftmargin=*, itemsep=2pt, topsep=3pt]
  \item \textbf{Quantitative noise mapping.} Complete test and train accuracy is reported for every combination of architecture, noise type, and error probability, establishing architecture-specific noise-tolerance bounds.
  \item \textbf{Amplitude encoding superiority.} Amplitude-inspired encoding achieves 100\% test accuracy up to $p = 0.10$ across all three noise channels---a ten-fold improvement in noise tolerance over entangled alternatives.
  \item \textbf{Architecture-specific immunity.} Z-encoding is provably immune to Phase-Flip noise at any error rate via the commutation $[R_Z(\theta),Z] = 0$; entangled circuits suffer the severest degradation through correlated two-qubit error propagation.
  \item \textbf{Generalisation gap analysis.} Noise induces train-test gaps of up to 17.5\% in entangled variants while Amplitude variants maintain zero gap, providing an actionable diagnostic for NISQ model development.
\end{enumerate}

Rebentrost et al.\cite{rebentrost2014} showed that quantum amplitude estimation gives exponential speedup of SVMs on large datasets. Havl\'{\i}\v{c}ek et al.\cite{havlicek2019supervised} implemented the compute-uncompute kernel protocol on superconducting hardware and introduced the ZZFeatureMap used here. Schuld and Killoran\cite{schuld2019} placed quantum feature maps within the kernel-methods framework, clarifying when quantum models differ from classical ones. Huang et al.\cite{huang2021} identified the conditions under which quantum kernels outperform classical alternatives, finding that dataset geometry is the decisive factor. Our work investigates how noise corrupts the kernel matrices that these methods depend on, and how that corruption varies by circuit structure.

\section{Methods and Theoretical Framework}

\subsection{Quantum Kernel Formulation}

For a supervised binary classification problem with training data $\mathcal{D} = \{(\mathbf{x}_i, y_i)\}_{i=1}^{N}$, where $\mathbf{x}_i \in \mathbb{R}^d$ and $y_i \in \{-1,+1\}$, a quantum feature map $\phi: \mathbb{R}^d \rightarrow \mathcal{H}^{2^n}$ encodes classical inputs as $n$-qubit states:
\begin{equation}
|\phi(\mathbf{x})\rangle = U(\mathbf{x})|0\rangle^{\otimes n},
\label{eq:featuremap}
\end{equation}
where $U(\mathbf{x})$ is a unitary that is a function of the classical input data. The quantum kernel is the squared fidelity between two encoded states:
\begin{equation}
K(\mathbf{x}_i, \mathbf{x}_j) =
\left|\langle 0|^{\otimes n} U^\dagger(\mathbf{x}_i) U(\mathbf{x}_j) |0\rangle^{\otimes n}\right|^2.
\label{eq:kernel}
\end{equation}
Eq.~\eqref{eq:kernel} is evaluated via the compute-uncompute protocol: prepare $U(\mathbf{x}_i)|0\rangle^{\otimes n}$, apply $U^\dagger(\mathbf{x}_j)$, and record the probability of the all-zeros measurement outcome. The kernel matrix $K_{ij}$ is then passed to a classical SVM, separating quantum feature extraction from classical quadratic-program optimisation.

\subsection{Feature Map Architectures}

\begin{figure}[!ht]
\centering
\begin{subfigure}[b]{0.46\linewidth}
\centering
\scalebox{0.92}{%
\begin{quantikz}[column sep=3pt, row sep=6pt]
\lstick{$q_0$} & \gate{H} & \gate{R_Z(x_0)} & \gate{H} & \gate{R_Z(x_0)} & \qw \\
\lstick{$q_1$} & \gate{H} & \gate{R_Z(x_1)} & \gate{H} & \gate{R_Z(x_1)} & \qw
\end{quantikz}}
\caption{\small ZFeatureMap (reps=2)}
\end{subfigure}
\hfill
\begin{subfigure}[b]{0.46\linewidth}
\centering
\scalebox{0.92}{%
\begin{quantikz}[column sep=3pt, row sep=6pt]
\lstick{$q_0$} & \gate{H} & \gate{R_Z(x_0)} & \ctrl{1}      & \qw \\
\lstick{$q_1$} & \gate{H} & \gate{R_Z(x_1)} & \gate{R_{ZZ}} & \qw
\end{quantikz}}
\caption{\small ZZFeatureMap (linear)}
\end{subfigure}

\vspace{6pt}

\begin{subfigure}[b]{0.46\linewidth}
\centering
\scalebox{0.92}{%
\begin{quantikz}[column sep=3pt, row sep=6pt]
\lstick{$q_0$} & \gate{H} & \gate{R_Z} & \ctrl{1} & \gate{R_{ZZ}} & \qw \\
\lstick{$q_1$} & \gate{H} & \gate{R_Z} & \targ{}  & \gate{R_{ZZ}} & \qw
\end{quantikz}}
\caption{\small PauliFeatureMap (Z, ZZ)}
\end{subfigure}
\hfill
\begin{subfigure}[b]{0.46\linewidth}
\centering
\scalebox{0.92}{%
\begin{quantikz}[column sep=3pt, row sep=6pt]
\lstick{$q_0$} & \gate{R_Y(x_0)} & \ctrl{1} & \gate{R_Z(x_0)} & \ctrl{1} & \qw \\
\lstick{$q_1$} & \gate{R_Y(x_1)} & \targ{}  & \gate{R_Z(x_1)} & \targ{}  & \qw
\end{quantikz}}
\caption{\small Amplitude-inspired}
\end{subfigure}

\caption{\textbf{Quantum circuit architectures for the four feature maps.} \textbf{(a)}~ZFeatureMap: alternating Hadamard and $R_Z$ layers, no entanglement. \textbf{(b)}~ZZFeatureMap: $R_Z$ encoding followed by a two-qubit $R_{ZZ}$ gate. \textbf{(c)}~PauliFeatureMap: Z and ZZ Pauli strings with linear CNOT entanglement. \textbf{(d)}~Amplitude-inspired: interleaved $R_Y$--CNOT--$R_Z$ layers that encode information in both amplitude and phase simultaneously.}
\label{fig:circuits}
\end{figure}
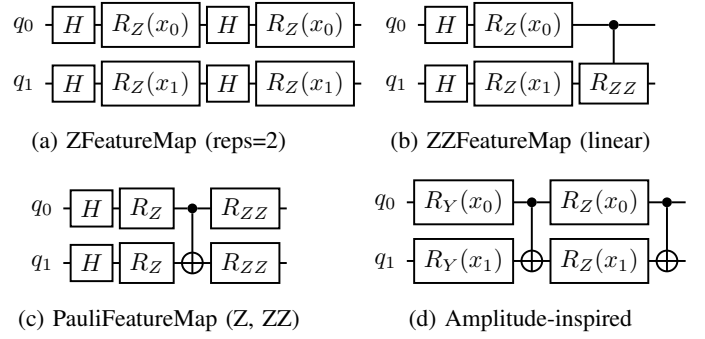

Four architectures with structurally distinct properties are evaluated. Figure~\ref{fig:circuits} shows the corresponding quantum circuits.

\paragraph{ZFeatureMap (Z)} Single-qubit Z-rotations alternate with Hadamard gates; no two-qubit gates appear: $U_Z(\mathbf{x}) = [H \cdot R_Z(x_i)]^{\mathrm{reps}}$. Shallow depth limits expressivity but also limits exposure to correlated noise.

\paragraph{ZZFeatureMap (ZZ)} ZFeatureMap is extended with two-qubit ZZ interactions $R_{ZZ}(\theta) = e^{-i\theta Z_iZ_j/2}$, where coupling angles encode feature correlations as $\phi_{ij}(\mathbf{x}) = (\pi - x_i)(\pi - x_j)$. This captures pairwise input dependencies at the cost of entangling gates.

\paragraph{PauliFeatureMap (Pauli)} Z and ZZ Pauli operator strings are combined with linear CNOT entanglement, giving flexible encoding at moderate circuit depth.

\paragraph{Amplitude-Inspired (Amplitude)} A custom architecture uses interleaved $R_Y$ and $R_Z$ rotation layers with strategically placed CNOT gates. Encoding data in both state amplitude ($R_Y$) and relative phase ($R_Z$) creates two partially independent pathways for preserving class information under complementary error types.

\subsection{Noise Models}

Three dominant decoherence mechanisms in NISQ hardware are modelled:

\paragraph{Depolarizing channel.} With probability $p$ the qubit state is replaced by the maximally mixed state, representing isotropic decoherence:
\begin{equation}
\mathcal{E}_{\mathrm{depol}}(\rho) = (1-p)\rho + \frac{p}{3}(X\rho X + Y\rho Y + Z\rho Z).
\end{equation}

\paragraph{Bit-Flip channel.} Pauli-X is applied with probability $p$, causing $|0\rangle \leftrightarrow |1\rangle$ transitions:
\begin{equation}
\mathcal{E}_{\mathrm{bit}}(\rho) = (1-p)\rho + pX\rho X.
\end{equation}

\paragraph{Phase-Flip channel.} Pauli-Z is applied with probability $p$, corrupting relative phases while preserving computational-basis populations:
\begin{equation}
\mathcal{E}_{\mathrm{phase}}(\rho) = (1-p)\rho + pZ\rho Z.
\end{equation}

Two-qubit gate errors are modelled as $\mathcal{E}^{(2)} = \mathcal{E} \otimes \mathcal{E}$, representing independent errors on each wire. Error probability was swept across $p \in \{0.01, 0.05, 0.10, 0.50\}$, spanning optimistic NISQ ($p = 0.01$), realistic NISQ ($p = 0.05$), high-noise ($p = 0.10$), and extreme decoherence ($p = 0.50$) regimes.

Let us note that $r1$, $r2$, etc. will be used in the paper to refer to the number of repetitions of encoding layers in the feature maps throughout the paper.

\subsection{Experimental Pipeline}

Algorithms~\ref{alg:noise} and~\ref{alg:qsvm} detail the noise model construction and QSVM evaluation pipeline.

\begin{algorithm}[!ht]
\caption{Noise Model Construction}
\label{alg:noise}
\begin{algorithmic}[1]
\REQUIRE Noise type $\tau \in \{\text{Depol.},\text{Bit-Flip},\text{Phase-Flip}\}$, error probability $p$
\ENSURE Qiskit \texttt{NoiseModel} $\mathcal{N}$
\STATE Initialise empty $\mathcal{N}$
\IF{$\tau = \text{Depolarizing}$}
  \STATE $\mathcal{E}_1 \leftarrow \texttt{depolarizing\_error}(p,1)$; $\mathcal{E}_2 \leftarrow \texttt{depolarizing\_error}(p,2)$
\ELSIF{$\tau = \text{Bit-Flip}$}
  \STATE $\mathcal{E}_1 \leftarrow \texttt{pauli\_error}([\text{X}:p,\,\text{I}:1{-}p])$; $\mathcal{E}_2 \leftarrow \mathcal{E}_1 \otimes \mathcal{E}_1$
\ELSIF{$\tau = \text{Phase-Flip}$}
  \STATE $\mathcal{E}_1 \leftarrow \texttt{pauli\_error}([\text{Z}:p,\,\text{I}:1{-}p])$; $\mathcal{E}_2 \leftarrow \mathcal{E}_1 \otimes \mathcal{E}_1$
\ENDIF
\STATE Attach $\mathcal{E}_1$ to all single-qubit gates; $\mathcal{E}_2$ to \texttt{cx}, \texttt{cz}
\RETURN $\mathcal{N}$
\end{algorithmic}
\end{algorithm}

\begin{algorithm}[!ht]
\caption{QSVM Training and Evaluation Pipeline}
\label{alg:qsvm}
\begin{algorithmic}[1]
\REQUIRE $(\mathbf{X}_\mathrm{tr},\mathbf{y}_\mathrm{tr})$, $(\mathbf{X}_\mathrm{te},\mathbf{y}_\mathrm{te})$, feature map $U$, noise type $\tau$, probability $p$
\ENSURE Train accuracy, test accuracy, training time
\IF{$\tau = \text{None}$}
  \STATE Use \texttt{StatevectorSampler} (exact simulation)
\ELSE
  \STATE $\mathcal{N} \leftarrow \textsc{NoiseModel}(\tau,p)$; initialise \texttt{AerSimulator}($\mathcal{N}$)
\ENDIF
\FORALL{pairs $(\mathbf{x}_i,\mathbf{x}_j)$}
  \STATE Build fidelity circuit: $C_\mathrm{fid} \leftarrow U(\mathbf{x}_i)\cdot[U(\mathbf{x}_j)]^\dagger$
  \STATE Execute with 1024 shots; $K_{ij} \leftarrow P(|0\rangle^{\otimes n})$
\ENDFOR
\STATE Train \texttt{SVC} with precomputed $K_\mathrm{tr}$; evaluate on $K_\mathrm{te}$
\RETURN Train accuracy, test accuracy, elapsed time
\end{algorithmic}
\end{algorithm}

\subsection{Dataset and Pre-processing}

The binary Iris dataset (Setosa vs.\ Versicolor, $N = 100$ samples) was used throughout. Sepal length and petal length were taken as features; both give reliable class separation and fit the two-qubit circuit architecture without requiring dimensionality reduction. All features were Z-score normalised (zero mean, unit variance) before encoding. A stratified 80/20 split with random seed 42 yielded 80 training samples (40 per class) and 20 test samples (10 per class). The full experiment matrix covers 4 feature maps $\times$ (1 noiseless baseline $+$ 3 noise types $\times$ 4 error probability levels) $=$ 52 controlled experiments, implemented in Qiskit~2.x with \texttt{qiskit-aer} for noise simulation and \texttt{scikit-learn} for the classical SVM.

\section{Results and Discussion}

\subsection{Baseline Performance}

A classical SVM with an RBF kernel achieved 100\% train and test accuracy, confirming that the two selected features fully separate Setosa from Versicolor. Under noiseless quantum simulation, test accuracies were 100\% for Amplitude, 90\% for ZZ, 85\% for Pauli, and 80\% for Z. Amplitude was the only encoding to match the classical ceiling; the other three left measurable accuracy unrealised despite mapping into an exponentially larger feature space than the RBF kernel uses.

\subsection{Noise Resilience: Main Findings}

\begin{figure*}[!ht]
\centering
\begin{subfigure}[b]{0.325\textwidth}
  \includegraphics[width=\linewidth]{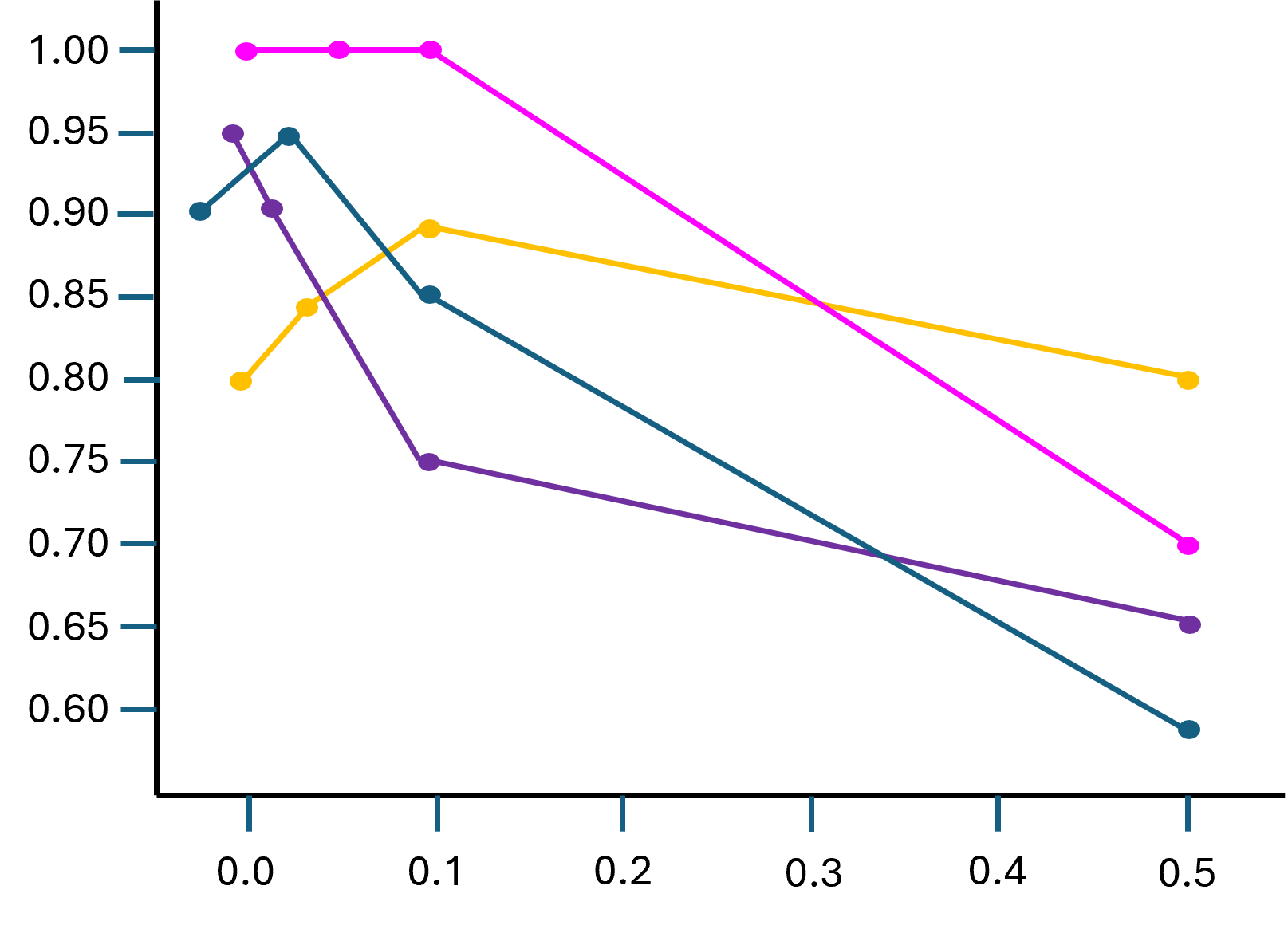}
  \caption{Depolarizing noise}
\end{subfigure}
\hfill
\begin{subfigure}[b]{0.325\textwidth}
  \includegraphics[width=\linewidth]{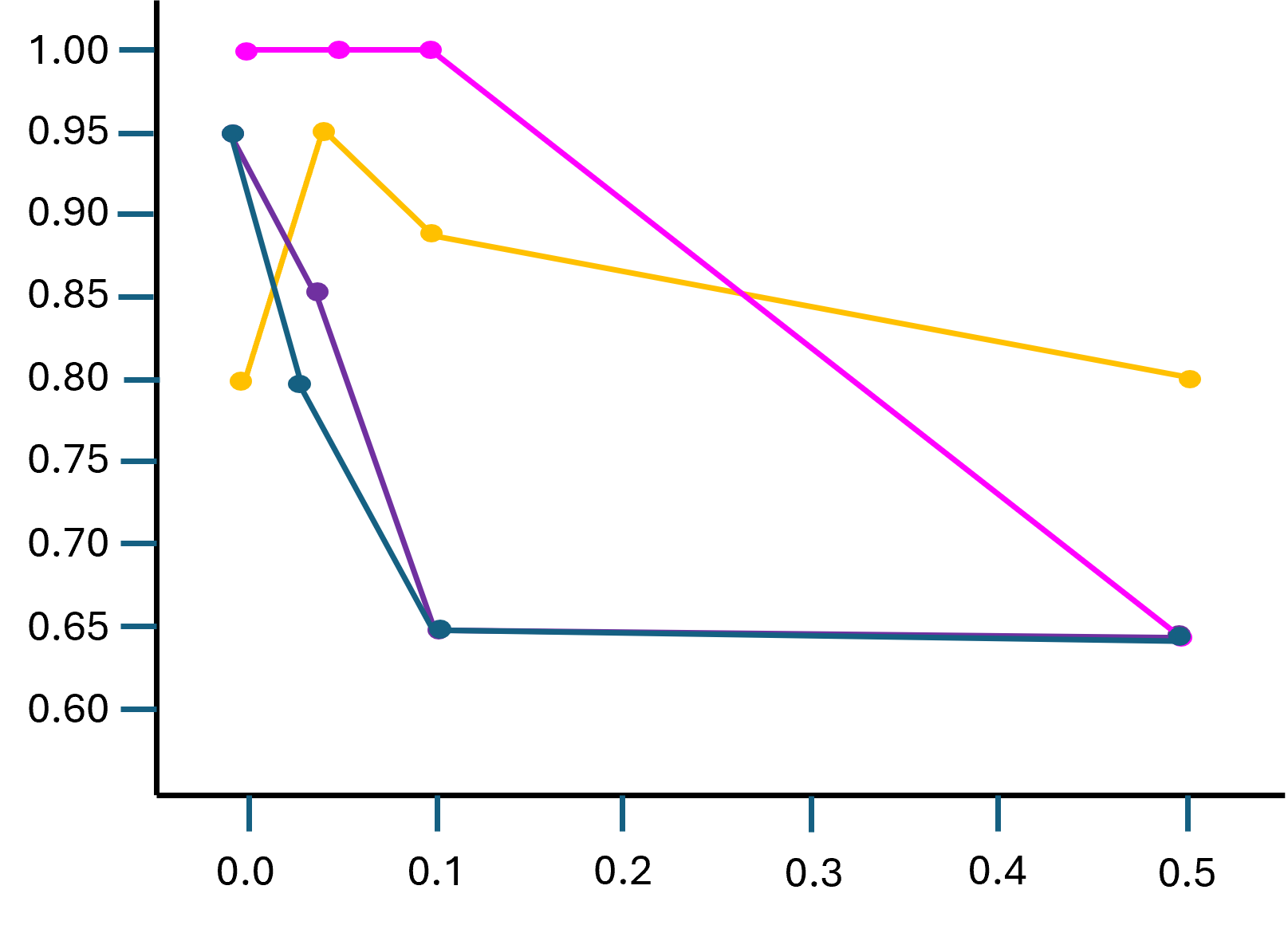}
  \caption{Bit-Flip noise}
\end{subfigure}
\hfill
\begin{subfigure}[b]{0.325\textwidth}
  \includegraphics[width=\linewidth]{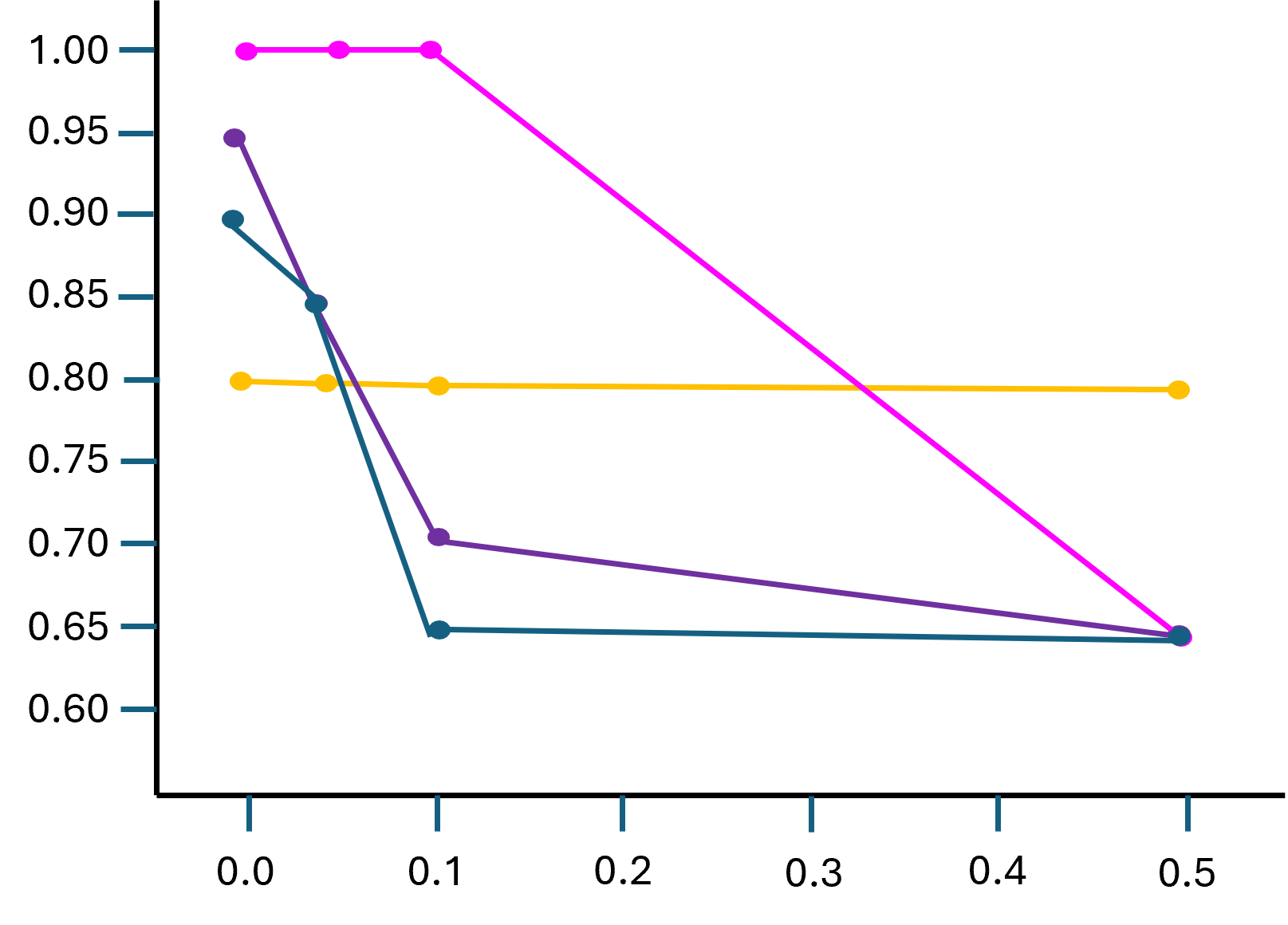}
  \caption{Phase-Flip noise}
\end{subfigure}
\caption{\textbf{Test accuracy versus noise probability for all three noise channels.} Each panel plots all four feature map architectures at $p \in \{0.01,0.05,0.10,0.50\}$. \textbf{(a) Depolarizing:} Amplitude (pink) sustains 100\% to $p = 0.10$ before declining; Z (yellow) rises initially via a regularisation effect before converging toward degradation. \textbf{(b) Bit-Flip:} Entangled architectures (ZZ, Pauli) degrade sharply at $p \geq 0.05$; Amplitude holds 100\% through $p = 0.10$. \textbf{(c) Phase-Flip:} Z-encoding (yellow) is perfectly flat at 80\% for all error rates, confirming Phase-Flip immunity via $[R_Z(\theta),Z]=0$. Amplitude again holds 100\% accuracy through $p = 0.10$ in this channel.}
\label{fig:noise_comparison}
\end{figure*}

Table~\ref{tab:main_results} compiles test accuracy across all 52 experiments. Figure~\ref{fig:noise_comparison} traces accuracy against $p$ for each architecture under each noise channel. Four principal findings emerge.

\begin{table}[!ht]
\centering
\caption{\textbf{Test accuracy (\%) under all experimental conditions.} Bold values indicate the best result per noise-probability column. Amplitude encoding sustains 100\% accuracy through $p = 0.10$ for every noise type. Entangled architectures (ZZ, Pauli) degrade to 60--65\% at $p = 0.50$.}
\label{tab:main_results}
\renewcommand{\arraystretch}{0.85}
{\small\setlength{\tabcolsep}{3pt}
\begin{tabular}{@{}llccccc@{}}
\toprule
\textbf{F. Map} & \textbf{Noise} & \textbf{$p{=}0.00$} & \textbf{$p{=}0.01$} & \textbf{$p{=}0.05$} & \textbf{$p{=}0.10$} & \textbf{$p{=}0.50$} \\
\midrule
\multirow{4}{*}{Z}
  & No Noise    & 80           & ---          & ---          & ---          & ---  \\
  & Depol.      & ---          & 80           & 85           & 90           & 80   \\
  & Bit-Flip    & ---          & 80           & 95           & 90           & 80   \\
  & Phase-Flip  & ---          & 80           & 80           & 80           & {80} \\
\midrule
\multirow{4}{*}{ZZ}
  & No Noise    & 90           & ---          & ---          & ---          & ---  \\
  & Depol.      & ---          & 90           & 95           & 85           & 60   \\
  & Bit-Flip    & ---          & 95           & 80           & 65           & 65   \\
  & Phase-Flip  & ---          & 90           & 85           & 65           & 65   \\
\midrule
\multirow{4}{*}{Pauli}
  & No Noise    & 85           & ---          & ---          & ---          & ---  \\
  & Depol.      & ---          & 95           & 90           & 75           & 65   \\
  & Bit-Flip    & ---          & 95           & 85           & 65           & 65   \\
  & Phase-Flip  & ---          & 95           & 85           & 70           & 65   \\
\midrule
\multirow{4}{*}{Amplitude}
  & No Noise    & {100} & ---          & ---          & ---          & ---  \\
  & Depol.      & ---          & {100} & {100} & {100} & 70 \\
  & Bit-Flip    & ---          & {100} & {100} & {100} & 65   \\
  & Phase-Flip  & ---          & {100} & {100} & {100} & 65   \\
\bottomrule
\end{tabular}
}
\end{table}

\paragraph{Finding 1: Amplitude encoding is robust to $p = 0.10$.}
Amplitude-inspired encoding held 100\% test accuracy at $p \in \{0.01, 0.05, 0.10\}$ for all three noise channels (Table~\ref{tab:main_results}). Accuracy dropped only at $p = 0.50$, where it reached 65--70\%. The usable error-rate window is ten times wider than that of entangled circuits: ZZ Bit-Flip accuracy, for example, drops from 95\% at $p = 0.01$ to 65\% at $p = 0.10$, while Amplitude holds 100\% across that range. The architecture encodes class information on two partially independent degrees of freedom: $R_Y$ rotations set state amplitudes, which Pauli-Z errors leave unchanged; $R_Z$ rotations set relative phases, which Pauli-X errors leave unchanged in the absence of subsequent Z-type gates. CNOT gates introduce inter-qubit correlations without adding the multi-layer entanglement that would compound errors exponentially with depth.

\paragraph{Finding 2: Z-encoding is immune to Phase-Flip noise.}
The Z feature map held exactly 80\% test accuracy at every Phase-Flip error rate from $p = 0.01$ to $p = 0.50$ (Figure~\ref{fig:noise_comparison}c). Every gate in the ZFeatureMap is either a Hadamard or an $R_Z$ rotation. Since $[R_Z(\theta), Z] = 0$ for all $\theta$, Phase-Flip errors (Pauli-Z applied with probability $p$) commute through the entire circuit without altering any measurement outcome. The same immunity does not extend to Bit-Flip errors; Pauli-X does not commute with $R_Z$ rotations, so Z-encoding remains vulnerable to that channel.

\paragraph{Finding 3: Mild noise improves accuracy in certain architectures.}
Several architectures scored higher at small $p > 0$ than under noiseless conditions:
\begin{itemize}[leftmargin=*, itemsep=0pt, topsep=2pt]
  \item Z under depolarizing: 80\% $\to$ 90\% at $p = 0.10$
  \item Z under bit-flip: 80\% $\to$ 95\% at $p = 0.05$
  \item ZZ under depolarizing: 90\% $\to$ 95\% at $p = 0.05$
  \item ZZ under bit-flip: 90\% $\to$ 95\% at $p = 0.01$
  \item Pauli under all noise types: 85\% $\to$ 95\% at $p = 0.01$
\end{itemize}
Random perturbations in kernel entries function as an implicit regulariser\cite{cerezo2021}: low-level noise disrupts the fine-grained kernel structure the SVM would otherwise overfit, producing decision boundaries that transfer better to held-out data. Amplitude does not exhibit this effect because its kernel entries remain accurate without it.

\paragraph{Finding 4: Entangled architectures degrade severely at high error rates.}
ZZ test accuracy fell to 60--65\% at $p = 0.50$ and Pauli to 65\%, both representing 25--30 percentage point losses from their noiseless values. A single fault on qubit $k$ before a CNOT$_{k \to l}$ exits as a correlated error on both $k$ and $l$; after $L$ entangling layers, that one initial fault can produce up to $2^L$ downstream errors. Nominal per-gate budgets therefore understate the actual noise exposure in ZZ and Pauli circuits, and the compounding grows with depth---which explains why the accuracy collapse at $p = 0.50$ is steeper than a simple per-gate model would predict.

\subsection{Training Accuracy and Noise-Induced Overfitting}
Table~\ref{tab:train_results} records training accuracy across all conditions. For Z (Depolarizing and Bit-Flip channels), ZZ, and Pauli, training accuracy climbed toward 100\% as $p$ rose while test accuracy fell in parallel. Noise warps the training kernel matrix in a way that makes training points appear more separable; the SVM fits that artefact and returns a boundary that does not hold on unseen data. Z Phase-Flip breaks this pattern: training accuracy stayed near 91\% at every error rate, because $[R_Z(\theta), Z] = 0$ stops Phase-Flip noise from reaching the kernel matrix at all (Finding~2). Amplitude showed no train-test divergence at any noise level or channel; both metrics held at 100\% throughout.

\begin{table}[!ht]
\centering
\caption{\textbf{Training accuracy (\%) under all experimental conditions.} Training accuracy rises toward 100\% with increasing noise for Z (Depolarizing and Bit-Flip), ZZ, and Pauli architectures while test accuracy simultaneously falls, confirming noise-induced overfitting. Z Phase-Flip is the single exception, remaining stable due to the commutation $[R_Z(\theta),Z]=0$. Amplitude encoding maintains 100\% train accuracy with no train-test divergence.}
\label{tab:train_results}
\renewcommand{\arraystretch}{0.85}
{\small\setlength{\tabcolsep}{3pt}
\begin{tabular}{@{}llccccc@{}}
\toprule
\textbf{F. Map} & \textbf{Noise} & \textbf{$p{=}0.00$} & \textbf{$p{=}0.01$} & \textbf{$p{=}0.05$} & \textbf{$p{=}0.10$} & \textbf{$p{=}0.50$} \\
\midrule
\multirow{4}{*}{Z}
  & No Noise    & 91.2  & ---    & ---    & ---    & ---    \\
  & Depol.      & ---    & 91.2  & 93.8  & 96.2  & 100.0 \\
  & Bit-Flip    & ---    & 92.5  & 95.0  & 100.0 & 100.0 \\
  & Phase-Flip  & ---    & 91.2  & 91.2  & 90.0  & 91.2  \\
\midrule
\multirow{4}{*}{ZZ}
  & No Noise    & 82.5  & ---    & ---    & ---    & ---    \\
  & Depol.      & ---    & 88.8  & 98.8  & 100.0 & 100.0 \\
  & Bit-Flip    & ---    & 86.2  & 98.8  & 100.0 & 100.0 \\
  & Phase-Flip  & ---    & 90.0  & 100.0 & 100.0 & 100.0 \\
\midrule
\multirow{4}{*}{Pauli}
  & No Noise    & 81.2  & ---    & ---    & ---    & ---    \\
  & Depol.      & ---    & 85.0  & 97.5  & 100.0 & 100.0 \\
  & Bit-Flip    & ---    & 90.0  & 100.0 & 100.0 & 100.0 \\
  & Phase-Flip  & ---    & 91.2  & 100.0 & 100.0 & 100.0 \\
\midrule
\multirow{4}{*}{Amplitude}
  & No Noise    & 100.0 & ---    & ---    & ---    & ---    \\
  & Depol.      & ---    & 100.0 & 100.0 & 100.0 & 100.0 \\
  & Bit-Flip    & ---    & 100.0 & 100.0 & 100.0 & 100.0 \\
  & Phase-Flip  & ---    & 100.0 & 100.0 & 100.0 & 100.0 \\
\bottomrule
\end{tabular}
}
\end{table}

\subsection{Circuit Variant Analysis}

\begin{figure*}[!ht]
\centering
\begin{subfigure}[b]{\textwidth}
  \includegraphics[width=\linewidth]{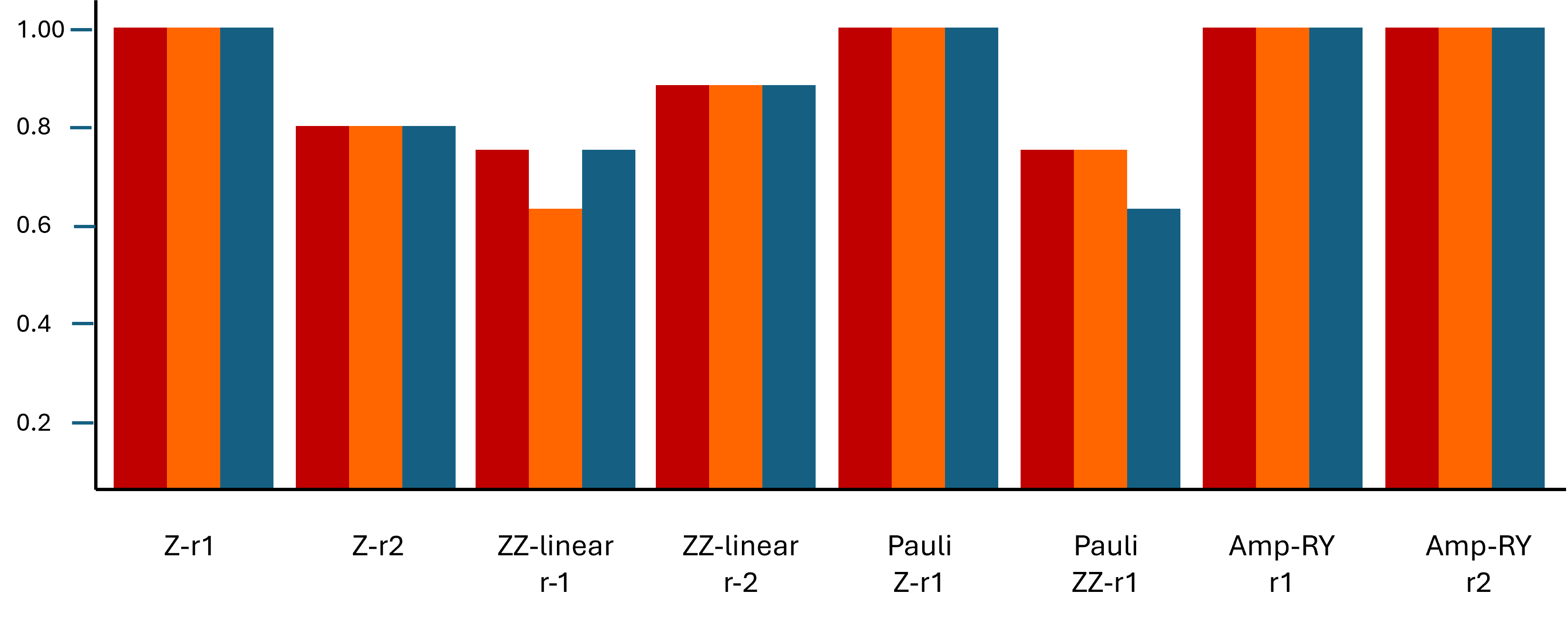}
  \caption{Test accuracy by variant and noise type}
\end{subfigure}

\vspace{6pt}

\begin{subfigure}[b]{\textwidth}
  \includegraphics[width=\linewidth]{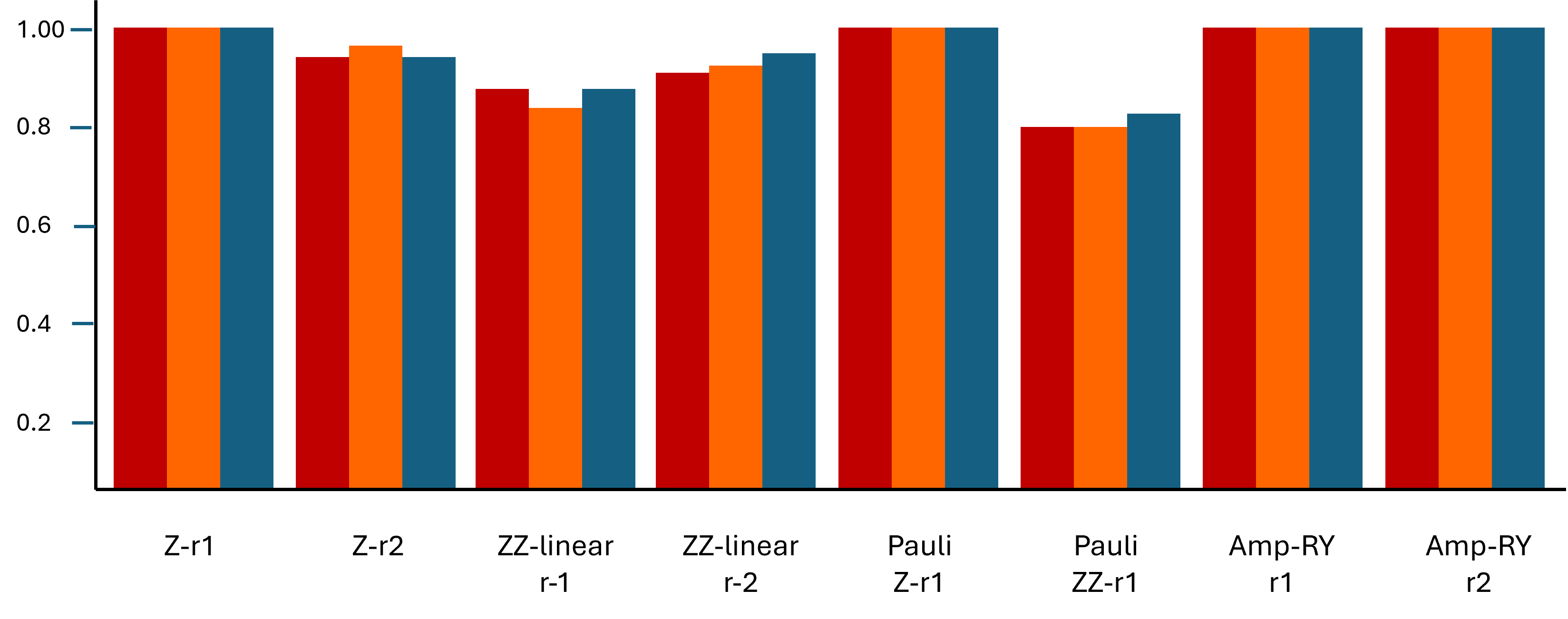}
  \caption{Training accuracy by variant and noise type}
\end{subfigure}
\caption{\textbf{Feature map variant performance at $p = 0.01$.} \textbf{(a) Test accuracy:} Z-r1, Pauli-Z-r1, Amp-RY-r1, and Amp-RY-r2 achieve 100\% across all four noise conditions. ZZ-lin-r1 drops to 65\% under Bit-Flip noise---the worst result across the entire study. ZZ-lin-r2 benefits from additional depth and achieves 90\% throughout. \textbf{(b) Training accuracy:} Variants with poor test accuracy simultaneously maintain near-100\% training accuracy, confirming noise-induced overfitting. Amp-RY variants hold 100\% train accuracy with zero generalisation gap, demonstrating that their robustness is structural.}
\label{fig:variants}
\end{figure*}

Eight circuit variants were evaluated at $p = 0.01$ to separate the contributions of circuit depth, entanglement structure, and rotation type. Table~\ref{tab:variants} reports the results; Figure~\ref{fig:variants} shows the accuracy bar charts.

\begin{table}[!ht]
\centering
\caption{\textbf{Circuit variant results at $p = 0.01$.} Train (Tr) and test (Te) accuracy (\%) for each variant and noise condition. Four variants achieve 100\% accuracy everywhere; entangled variants ZZ-lin-r1 and Pauli-ZZ-r1 show the largest drops and the highest generalization gaps.}
\label{tab:variants}
\renewcommand{\arraystretch}{0.9}
{\small\setlength{\tabcolsep}{3pt}
\begin{tabular}{@{}lcccccccc@{}}
\toprule
& \multicolumn{2}{c}{\textbf{No Noise}} & \multicolumn{2}{c}{\textbf{Depolarizing}} & \multicolumn{2}{c}{\textbf{Bit-Flip}} & \multicolumn{2}{c}{\textbf{Phase-Flip}} \\
\cmidrule(lr){2-3}\cmidrule(lr){4-5}\cmidrule(lr){6-7}\cmidrule(lr){8-9}
\textbf{F. Map} & Tr & Te & Tr & Te & Tr & Te & Tr & Te \\
\midrule
Z-r1          & 100.0 & 100.0 & 100.0 & 100.0 & 100.0 & 100.0 & 100.0 & 100.0 \\
Z-r2          & 91.2 & 80.0 & 91.2 & 80.00 & 92.5 & 80.0 & 91.2 & 80.0 \\
ZZ-lin-r1     & 80.0 & 70.0 & 83.8 & 75.00 & 82.5 & 65.0 & 83.8 & 75.0 \\
ZZ-lin-r2     & 81.2 & 90.0 & 85.0 & 90.00 & 91.2 & 90.0 & 93.8 & 90.0 \\
Pauli-Z-r1    & 100.0 & 100.0 & 100.0 & 100.0 & 100.0 & 100.0 & 100.0 & 100.0 \\
Pauli-ZZ-r1   & 81.2 & 75.0 & 81.2 & 75.00 & 81.2 & 75.0 & 82.5 & 65.0 \\
Amp-RY-r1     & 100.0 & 100.0 & 100.0 & 100.0 & 100.0 & 100.0 & 100.0 & 100.0 \\
Amp-RY-r2     & 100.0 & 100.0 & 100.0 & 100.0 & 100.0 & 100.0 & 100.0 & 100.0 \\
\bottomrule
\end{tabular}
}
\end{table}

Three structural patterns emerge. Variants using only single-qubit rotations (Z-r1, Pauli-Z-r1) or the dual-rotation Amplitude scheme (Amp-RY-r1, Amp-RY-r2) all reached 100\% accuracy, putting entanglement depth as the primary noise risk factor. ZZ-lin-r1 (one repetition, linear entanglement) hit 65\% test accuracy under Bit-Flip---the lowest result in this study. Depth interacts non-monotonically with architecture: an extra repetition lifts ZZ test accuracy from 70\% to 90\% at the noiseless baseline (ZZ-lin-r1 vs.\ ZZ-lin-r2), but drops Z from 100\% to 80\% (Z-r1 vs.\ Z-r2), because additional Z-rotation layers narrow the noise tolerance that the shallower variant preserves.

\subsection{Generalisation Gap Analysis}

Table~\ref{tab:gen_gap} quantifies the train-test accuracy gap (Train $-$ Test) for all variants at $p = 0.01$.

\begin{table}[!ht]
\centering
\caption{\textbf{Generalisation gap at $p = 0.01$ (Train $-$ Test, \%).} Green cells: zero gap (perfect generalisation). Red cells: severe overfitting ($\geq$15\%). Two entangled variants produce critical 17.5\% gaps; four variants maintain zero gap across all conditions.}
\label{tab:gen_gap}
\renewcommand{\arraystretch}{1.05}
{\small\setlength{\tabcolsep}{4.5pt}
\begin{tabular}{@{}lcccc@{}}
\toprule
\textbf{F. Map} & \textbf{No Noise} & \textbf{Depolarizing} & \textbf{Bit-Flip} & \textbf{Phase-Flip} \\
\midrule
Z-r1          & \cellcolor{bestcolor}0.0  & \cellcolor{bestcolor}0.0 & \cellcolor{bestcolor}0.0          & \cellcolor{bestcolor}0.0  \\
Z-r2          & 11.2                      & 11.2                     & 12.5                              & 11.2                      \\
ZZ-lin-r1     & 10.0                      & 8.8                      & \cellcolor{worstcolor}17.5        & 8.8                       \\
ZZ-lin-r2     & $-$8.8                    & $-$5.0                   & 1.2                               & 3.8                       \\
Pauli-Z-r1    & \cellcolor{bestcolor}0.0  & \cellcolor{bestcolor}0.0 & \cellcolor{bestcolor}0.0          & \cellcolor{bestcolor}0.0  \\
Pauli-ZZ-r1   & 6.2                       & 6.2                      & 6.2                               & \cellcolor{worstcolor}17.5 \\
Amp-RY-r1     & \cellcolor{bestcolor}0.0  & \cellcolor{bestcolor}0.0 & \cellcolor{bestcolor}0.0          & \cellcolor{bestcolor}0.0  \\
Amp-RY-r2     & \cellcolor{bestcolor}0.0  & \cellcolor{bestcolor}0.0 & \cellcolor{bestcolor}0.0          & \cellcolor{bestcolor}0.0  \\
\bottomrule
\end{tabular}
}
\end{table}

ZZ-lin-r1 under Bit-Flip and Pauli-ZZ-r1 under Phase-Flip both produced 17.5\% gaps (Train~=~82.5\%, Test~=~65.0\%). In both cases the classifier memorised the training kernel matrix well but transferred poorly: the 82.5\% training accuracy, rather than 100\%, shows that noise also perturbs training entries, yet the perturbation pattern on training data differs enough from that on test data to produce a 17.5 percentage-point gap. Four variants---Z-r1, Pauli-Z-r1, Amp-RY-r1, Amp-RY-r2---maintained zero gap across every condition. ZZ-lin-r2 showed a $-$8.75\% gap under no noise, where test accuracy exceeded training accuracy; this reflects sampling variation in a 20-sample test set rather than any structural generalization advantage.

\subsection{Aggregate Performance Summary}

\begin{figure*}[!ht]
\centering
\begin{subfigure}[b]{0.49\textwidth}
  \includegraphics[width=\linewidth]{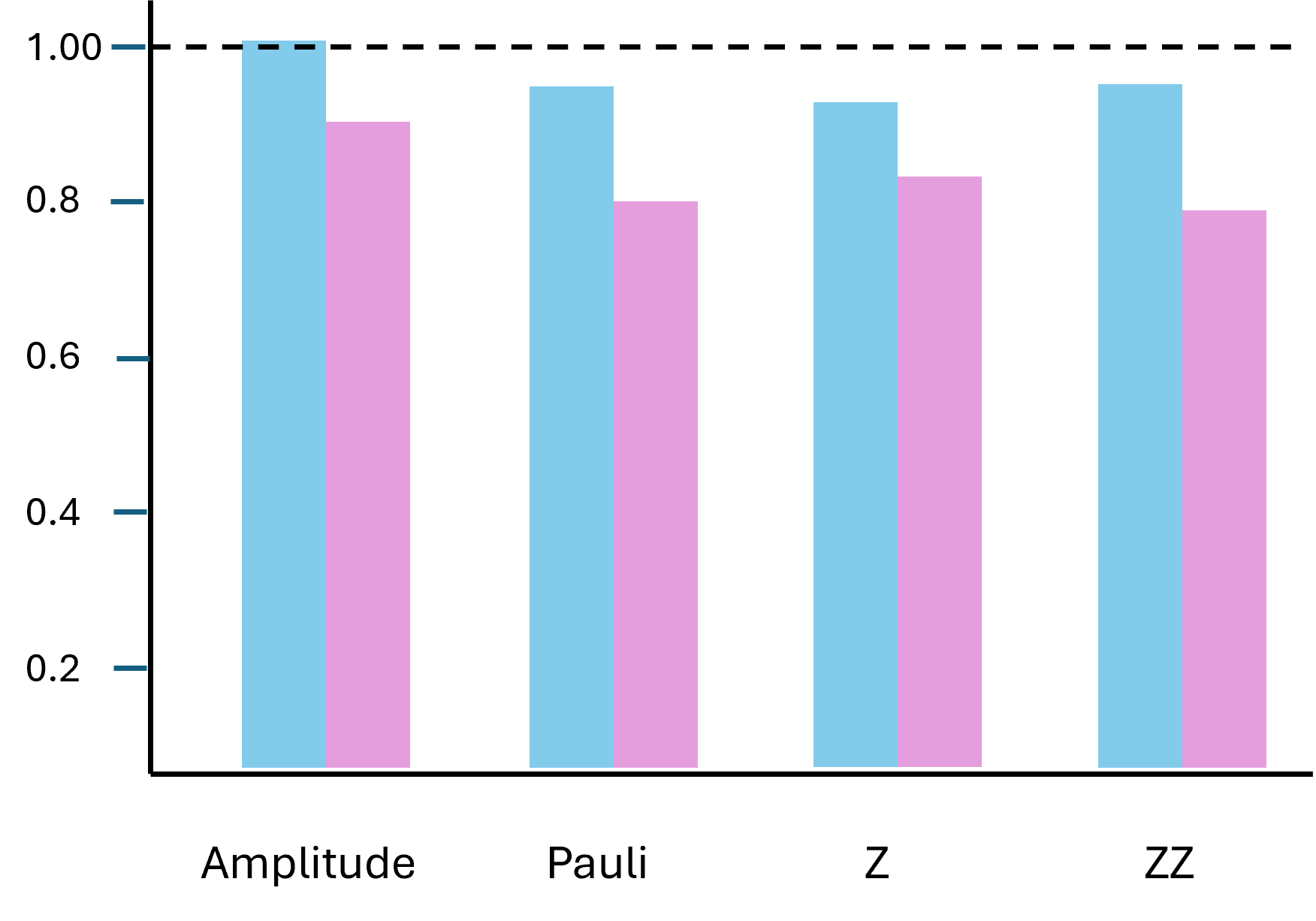}
  \caption{Average train and test accuracy by feature map}
\end{subfigure}
\hfill
\begin{subfigure}[b]{0.49\textwidth}
  \includegraphics[width=\linewidth]{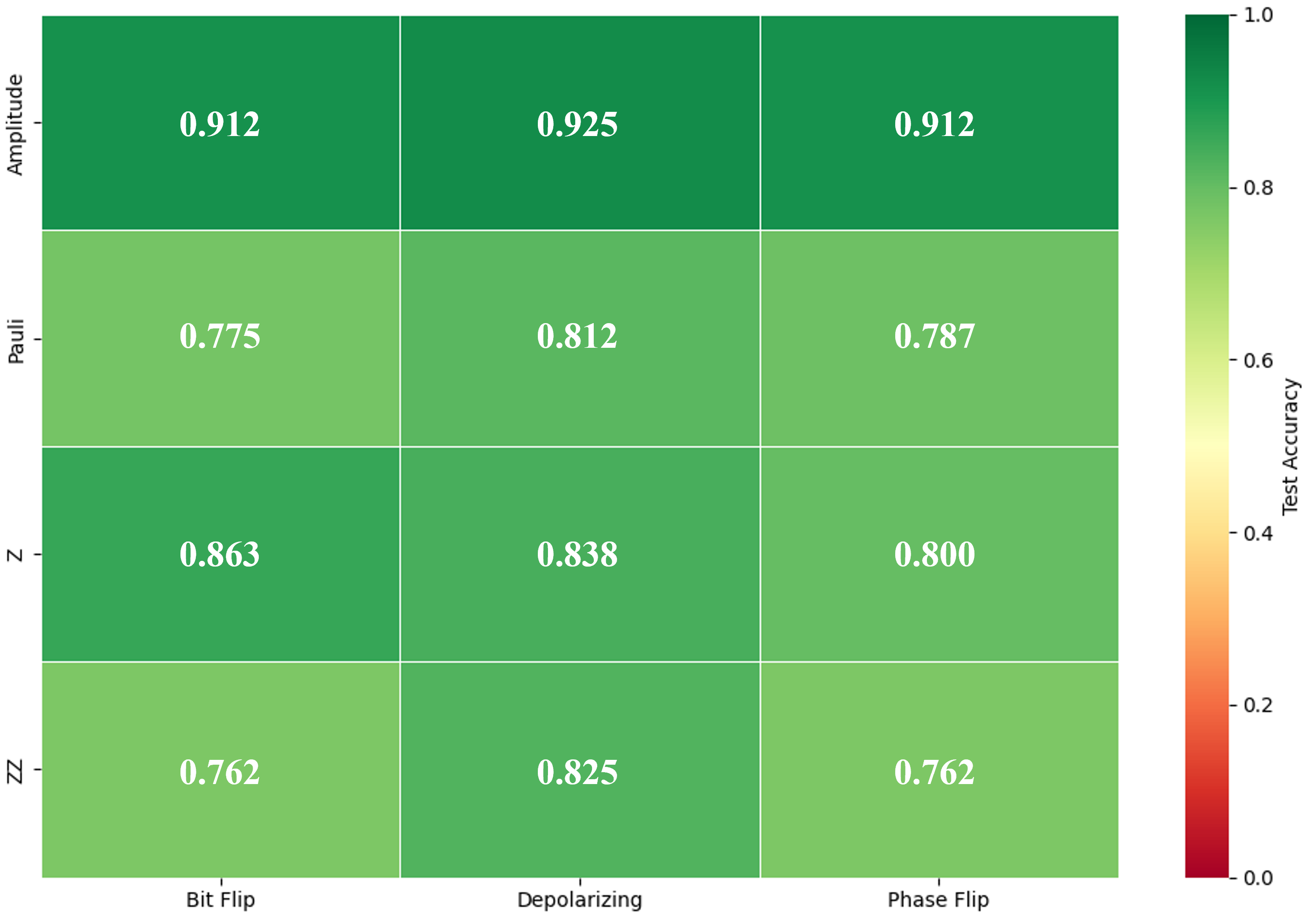}
  \caption{Average test accuracy heatmap}
\end{subfigure}
\caption{\textbf{Aggregate performance across all noise conditions.} \textbf{(a)} Mean train (blue) and test (pink) accuracy for each feature map, averaged over all noise types and error probability levels. Amplitude achieves the highest test accuracy with the smallest train-test gap; ZZ shows the largest divergence, confirming systematic noise-induced overfitting. (b) Average test accuracy heatmap. Cell values report the mean test accuracy across $p \in \{0.01, 0.05, 0.10, 0.50\}$ for each feature-map–noise-type combination. Amplitude (top row) exceeds 0.91 for all channels; ZZ (bottom row) drops to 0.762 under Bit-Flip and Phase-Flip, the clearest evidence of entanglement-mediated noise vulnerability in this study.}
\label{fig:summary}
\end{figure*}

Averaged across all noise types and error probability levels, Amplitude encoding reached 91--93\% mean test accuracy with a near-zero train-test gap (Figure~\ref{fig:summary}). Z-encoding ranked second at 80--86\%, held from falling further by its Phase-Flip immunity. ZZ averaged 76.2\% under Bit-Flip and Phase-Flip; Pauli averaged 77.5\% under Bit-Flip. The heatmap in Figure~\ref{fig:summary}b makes this ranking explicit: Amplitude's row stays above 0.91 across all three channels, while ZZ drops to 0.762 under Bit-Flip and Phase-Flip.

\subsection{Mechanistic Interpretation}

\paragraph{Why Amplitude encoding resists all noise types.}
$R_Y$ rotations set state amplitudes; Pauli-Z errors (Phase-Flip, and the Z-component of Depolarizing) leave amplitudes unchanged. $R_Z$ rotations set relative phases; Pauli-X errors (Bit-Flip, and the X-component of Depolarizing) leave relative phases unchanged in circuits without further Z-type gates. The RY--CNOT--RZ structure therefore keeps class information partially intact under both Pauli-X and Pauli-Z perturbations independently. CNOT gates couple the two qubits without adding additional entangling layers that would amplify single-qubit errors into multi-qubit correlated errors across repetitions.

\paragraph{Why Z-encoding is immune to Phase-Flip.}
Phase-Flip errors apply Pauli-Z with probability $p$. Since $[R_Z(\theta), Z] = 0$ for all $\theta$, inserting a Pauli-Z error at any point in a circuit consisting entirely of Hadamard and $R_Z$ gates is equivalent to inserting it at a different point, without altering the final state or any measurement probability. Phase-Flip errors permute the temporal order of operations; that permutation leaves the circuit output invariant.

\paragraph{Why entanglement amplifies noise.}
A single-qubit error on qubit $k$ before a CNOT$_{k\to l}$ gate becomes a correlated two-qubit error on both $k$ and $l$ after the gate. In a circuit with $L$ entangling layers, one initial error can produce up to $2^L$ correlated downstream errors, making the effective error rate grow super-linearly with circuit depth. This accounts for the 25--30 percentage point degradation of ZZ and Pauli at $p = 0.50$.

\subsection{Practical Deployment Guidelines}

\paragraph{Architecture selection.}
\begin{enumerate}[leftmargin=*, itemsep=2pt, topsep=4pt]
  \item \textit{Primary recommendation.} Use Amplitude-based encoding (Amp-RY-r1 or Amp-RY-r2) for hardware with per-gate error rates $p \leq 0.10$. Both variants delivered 100\% test accuracy across all tested noise types with zero generalisation gap, covering the full realistic NISQ operating range ($p \approx 0.05$).
  \item \textit{Secondary recommendation.} Use Z-r1 or Pauli-Z-r1 when Phase-Flip is the dominant noise mechanism or when hardware error rates satisfy $p \leq 0.05$. Both achieve 100\% accuracy at $p = 0.01$ and are provably immune to Phase-Flip noise at any rate.
  \item \textit{Architectures to avoid.} ZZ-lin-r1 and Pauli-ZZ-r1 produced the lowest test accuracy (65\%) and the largest generalisation gaps (17.5\%) under NISQ conditions. Dedicated error mitigation is needed before either reaches hardware.
\end{enumerate}

\paragraph{Diagnostic monitoring.}
A train-test gap above 10\% during model development signals that noise has corrupted the kernel matrix enough to cause overfitting. Switching to a shallower or Amplitude-based architecture is the appropriate response before hardware deployment.

\subsection{Limitations}

\paragraph{Dataset scope.}
All experiments used 100 samples, 2 features, and 2 qubits. The two-class Iris problem is too simple to stress-test the expressivity of larger feature maps; the Amplitude architecture may owe part of its advantage to the classification task's simplicity rather than to noise resilience alone. Experiments on 4--10 qubit circuits with datasets where quantum kernels offer genuine representational advantage over classical methods are needed before these conclusions extend beyond 2-qubit systems.

\paragraph{Idealised noise model.}
Spatially and temporally uniform Pauli error channels with a fixed per-gate probability were applied throughout. Real NISQ hardware has gate-dependent error rates, qubit-specific decoherence times, crosstalk between neighbouring qubits, leakage to non-computational states, and time-varying drift. Architecture selection conclusions from this study need validation against device-specific noise profiles before application to particular hardware platforms.

\paragraph{Simulation environment.}
All experiments ran on \texttt{qiskit-aer} noise simulation, not physical hardware. Validation on IBM Quantum, IonQ, or Rigetti processors is required to confirm that the noise-resilience ordering observed in simulation holds under real decoherence.

\section{Conclusions}

Amplitude-inspired feature map held 100\% test accuracy through $p = 0.10$ across all three noise channels and produced zero train-test generalization gap under every tested condition. The Z feature map is immune to phase-flip noise at high error rate, a direct consequence of the commutation $[R_Z(\theta), Z] = 0$. Entangled circuits produced generalization gaps of up to $17.5$\% and fell to $60$-$65$\% accuracy at $p = 0.50$. Mild depolarizing and bit-flip noise improved accuracy by $5$-$15$ percentage points for feature maps other than the amplitude-inspired through a kernel-level regularization mechanism. For the noisy intermediate scale quantum (NISQ) device users, these results supply a ranked architecture guide: amplitude-based encoding for hardware up to $p = 0.10$; single-qubit rotation maps when phase-flip dominates; and entangled circuits only when dedicated error mitigation is available.

Extending this study to larger qubit systems and real hardware is a clear next step. Experiments on $4$–$8$ qubit amplitude-inspired feature maps with datasets where classical kernels cannot match quantum expressivity would test whether the noise-resilience advantage scales, while device-specific noise profiles from hardware calibration data would clarify how well the simulation-derived hierarchy transfers to physical processors.


\section*{Author Contributions}
M.A.S. implemented the quantum algorithms and conducted all experiments, S.M. developed the noise analysis framework and produced visualizations, D.T. performed statistical analysis and classical baseline comparisons, and M.F. conceived the study, designed the experimental framework, and supervised all aspects of the work, with all authors contributing to data interpretation and manuscript preparation.

\section*{Competing Interests}
The authors declare no competing financial or non-financial interests.

\end{document}